\documentclass[12pt]{article}
\usepackage{amsmath}
\usepackage{amssymb}
\usepackage{url}
\usepackage{setspace}
\usepackage{bm}
\begin{document}
\onehalfspacing
\newpage \pagenumbering{arabic}
\medskip
\begin{center}
\LARGE{A Possible Mathematical Structure for Physics}
\end{center}
\medskip
\begin{center}
Lester C. Welch\footnote{Email: lester.welch@gmail.com }

Aiken, SC 29803

\end{center}
\begin{quote}
The relationship between mathematics and physics has long been an area of interest and speculation.  Subscribing to the recent definition by Tegmark, we present a mathematical structure involving the only division rings - the real, $\mathbb{R}$, the complex, $\mathbb{C}$, and the quaternion, $\mathbb{H}$ - known to meet the requirements of a mathematical description of our universe.   The known forces and particle structure, in terms of families and color, can be associated with natural relationships within that structure.   A "toy" scientific theory - based on the proposed mathematical structure - is presented to illustrate its applicability.   It is shown how parity and time reversal invariance are broken. Two relevant predictions come from the formalism. First, there are no gravity waves and second, matter comes in three varieties, one of which is "dark" in the sense that it is oblivious to electromagnetic interactions.
\end{quote}
\section{Background}

The relationship between mathematics and physics has been a matter of study since the time of the Pythagoreans.  Notables such as Galilei, Hawking, Wheeler, and Wigner have pondered - to paraphrase each - "Why is mathematics so successful in describing the universe?"  Recently Tegmark \cite{Teg1} has done an in depth study involving the tools of philosophy, mathematical descriptions, and physical systems to explore the relationship and to identify the requirements of a mathematical structure - such as computations and simulations - to represent physical reality and the reader is referred to his exhaustive set of references.   Tegmark argues that \textit{"Our external physical reality is a mathematical universe"} - the Mathematical Universe Hypothesis (MUH).  Following his description Benioff \cite{Ben1} began the development of a complete framework for a coherent theory of physics and mathematics involving a space and time lattice where each frame has finite qukit strings.  Bernal, \textit{et.al.}\cite{Bernal} investigated the symmetries of spacetime in the guise of a Tegmark mathematical structure and concluded that only four types of spacetime are possible. \\ \\
This work is in that spirit. I offer a mathematical structure that fits the Tegmark prescription, but is equipped with sufficient details to form physical theories.  In a sense the current work is intermediate between Tegmark's work and a scientific theory. Stretching Tegmark's terminology, the view is from a bird that has descended and is at the top of a tall tree.\\ \\
A important point of Tegmark's, relevant to the present work, is his definition of a mathematical structure.  To quote:
\begin{quote}
\textit{
We define a mathematical structure as a set \textit{S} of abstract entities and relationships $R_1, R_2,...$ between them.  Specifically, let us define the entities and relations of a mathematical structure as follows: \\ \\
Given a finite number of sets $S_1, S_2,...S_n$,\\ \\
1. the set of entities is the union $S=S_1\cup S_2\cup...\cup S_n$,\\ \\
2. the relationships are functions on these sets, specifically mappings from some number of sets to a set: $S_{i_1}\times S_{i_2}\times ...S_{i_k}\mapsto S_j$.}
\end{quote}

\section{The Mathematical Structure}
Birkhoff and Neumann\cite{Birk1} formulated a set of axioms obeyed by quantum mechanics - a "propositional calculus", which is roughly speaking equivalent to the Hilbert space approach.  Recently, Aerts\cite{aerts} has reviewed the historical development and the present status of Birkhoff and Neumann's approach.  Because of the Frobenius theorem, there are only three mathematical fields (division rings),  the real, $\mathbb{R}$, the complex, $\mathbb{C}$, and the quaternion, $\mathbb{H}$ from which one can construct a quantum mechanics fulfilling the propositional calculus.  The field, perhaps less familiar to the reader, is the quaternion.  The quaternions can be roughly characterized as an extension, involving three non-commuting imaginaries\footnote{The quaternion imaginaries will be labeled $I,J,K$ to avoid confusion with $i\in \mathbb{C}$.  They obey the algebraic rules: $II=JJ=KK=-1$ and $IJ=K$, $JK=I$, and $KI=J$.} similar to the familiar complex field.  Since their discovery by Hamilton in 1843 and the pioneering work of Finkelstein, $\emph{et.,al}$ \cite{Finkelstein2}, quaternions  have received a great deal of attention, e.g., \cite{Yefremov,Frenkel,Negi,Leo1,Khalek,Adler1,Maia,Rawat,Rawat2}, as a mathematical formalism for expressing physics.  The most comprehensive and notable study of their applicability in quantum mechanics has been done by Adler\cite{Adler}. However, their use has not been as universal as might be expected given the non-commutative nature of both quaternions and popular modern theories that involve non-Abelian guage invariance and higher order symmetries ($e.g., SU(3)$). In previous works\cite{welch, welch1}, I explored the use of quaternions in the solution of Dirac's equation and showed how color conservation and fractional charges could result because of the richness provided by an increased number of involutions in a manner not easily achieved in $\mathbb{C}$, however it became clear to me that a single field - complex or quaternion - was not likely to explain all features of reality and thus was lead to the current work.  As Aerts\cite{aerts} asked,
\begin{quote}
\textit{Why would a complex Hilbert space deliver the unique mathematical structure for a complete description of the microworld?  Would that not be amazing?  What is so special about a complex Hilbert space that its mathematical structure would play such a fundamental role?}
\end{quote}
\subsection{Structure Entities}
It is necessary to explain the viewpoint adopted herein.  The \textbf{MUH} is accepted \textit{in toto} and a mathematical structure is proposed that has three sub-structures (each describing a sub-universe) composed of $\mathbb{R}$ (with the element $1_R$), $\mathbb{C}$ (with elements, $1_C$ and $i$) and $\mathbb{H}$ (with elements, $1_H, I, J,$ and $K$).  By maintaining three separate sub-structures, the identity of the "real" element in each field, $1_R, 1_C$ and $1_H$ can be maintained and play a necessary role in the specification of the relationships needed to define the structure as a whole. One additional element is needed which is a \textit{c-number} - potentially a multiplier of each of the elements of $\mathbb{R}$, $\mathbb{C}$, and $\mathbb{H}$ and commutes with all of them.  In other words, the set $S$ of the structure can be thought of as the elements of $\mathbb{R} \times \mathbb{C} \times \mathbb{H}$ plus a \textit{c-number} and it would be wrong to view $\mathbb{R}$ as isomorphic to a subset of $\mathbb{C}$ or $\mathbb{H}$ or, similarly $\mathbb{C}$ as a isomorphic subset of $\mathbb{H}$.  In a sense there are three sub-universes which interact to form the observed universe.  One sub-universe, $U_R$, has $\mathbb{R}$ as its mathematical structure, one, $U_C$ has $\mathbb{C}$ as the mathematical structure and the third, $U_H$ has the elements of the quaternion field as its structure. In a sense $\psi=\psi_R\psi_C\psi_H$. As John Baez \cite{baez1} said: 
\begin{quote}
\textit{
...one can't help but think that complex, real, and quaternionic
quantum mechanics fit together in a unified structure, with the complex
numbers being the most important,...}  
\end{quote}
The view taken herein is that the entities of the mathematical structure are complete and known and that the role of physics is to discover the relationships, $R_i$, that complete the mathematical structure.  It is further hypothesized that each sub-universe - because of the elements of the field used in its mathematical structure - can only accommodate certain relationships. \\ \\  Gravity, the simplest of all of the interactions, is assumed - as an ansatz - to be manifested in $U_R$ where there is only one element, 1$_R$, whose multipliers (\textit{c-numbers}) can assume positive or negative values but no concept of a "phase" between elements.  With regard to its multipliers, it is one dimensional - sufficient to quantify the concept of "mass."  \\ \\The complex field, $\mathbb{C}$, however, has two commutating elements 1$_C$ and $i$.  A phase can be defined in the two dimensional space spanned by its elements.  This allows sufficient flexibility so that Maxwell's equations can be constructed to describe electromagnetic waves in $U_C$.  Quantum electrodynamics (QED) is the most successful scientific theory and uses only Abelian relationships using $\mathbb{C}$ within $U_C$, in the mathematical structure discussed herein .  It seems reasonable to assert that all non-hadronic electromagnetic relationships are described in $U_C$.\\ \\  Strong forces - quantum chromodynamics - however, need a richer structure to accommodate the additional relationship of "color." The three additional non-commutating elements, $I,J,K$, can accommodate "color" within $U_H$. The entities of the mathematical structure are depicted in \textit{Figure 1.}
\vspace{.43in}
\begin{center}\setlength{\unitlength}{1in}
\begin{picture}(0,0)
\put(-1.6,.2){\text Gravity, $U_R$}
\put(-.6,.2){\text Leptons, $U_C$}
\put(.4,.2){\text Hadrons, $U_H$}



\put(-1.3,-1.8){$\mathbb{R}$}
\put(-.3,-1.8){$\mathbb{C}$}
\put(.7,-1.8){$\mathbb{H}$}
\put(-.6,-.73){\em{i}}
\put(.32,-.53){\em{J}}
\put(.32,-.05){\em{K}}
\put(.32,-1.04){\em{I}}
\put(.32,-1.54){$\mathit{1}_H$}
\put(-1.7,-1.54){$\mathit{1}_R$}
\put(-.7,-1.54){$\mathit{1}_C$}
\linethickness{.03in}
\put(.5,0){\line(1,0){.4}}
\put(.5,-.5){\line(1,0){.4}}
\put(.5,-1){\line(1,0){.4}}
\put(-.5,-.7){\line(1,0){.4}}
\put(-1.5,-1.5){\line(1,0){.4}}
\put(-.5,-1.5){\line(1,0){.4}}
\put(.5,-1.5){\line(1,0){.4}}
\end{picture}            

\vspace{2in}
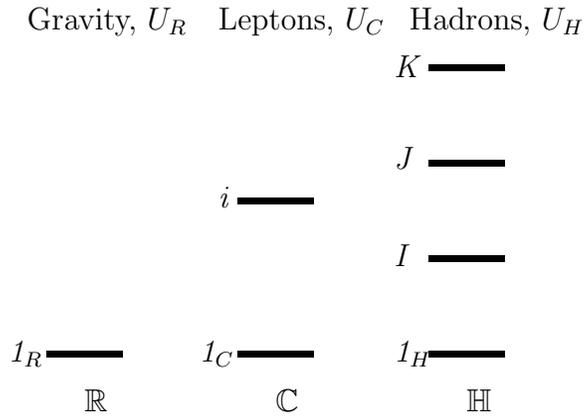
Figure 1.  The three mathematical rings suitable for physics whose elements are the entities for the mathematical structure discussed herein. Each of the elements can be multiplied by a \textit{c-number}.
\end{center}

\subsection{The Relationships}
The relationships among the entities create the physics.  Since $U_R$ is stipulated to be the sub-universe within which only gravity manifests itself, the sole element 1$_R$ therefore is in someway responsible for mass in $U_R$.  In a similar way, 1$_C$ is responsible for mass in $U_C$ and likewise, 1$_H$ in $U_H$.  These ideas will be made clearer in the "toy" theory developed later.\\ \\
Another ansatz is that $e^{\pm}$ are due to a relationship totally within $U_C$ since that is the lowest dimensional sub-universe in which an electromagnetism can be formulated.  The muon $\mu^{\pm}$ and tau $\tau^{\pm}$ leptons are also manifested in $U_C$ but there is an additional relationship - a "coupling" among 1$_R$, 1$_C$ and 1$_H$ - that causes a mass spectrum.   Thus a natural explanation of the existence of "families" is that there are couplings among the elements responsible for mass - the real numbers, 1$_R$, 1$_C$, and 1$_H$.  The "toy" theory illustrates how this is possible.\\ \\
When one considers the possible relationships within $U_H$ it is natural to ascribe the three colors - in some way - to the three imaginaries, $I,J,K.$  In \cite{welch1}, I showed how conjugating $I,J,K$ independently led to involutions which preserves three additional currents - which can be associated with "color" - much in the same way that complex conjugation in $U_C$ leads to conservation of electrical charge via a conserved current.  If one does a complete conjugation of $I,J,K$ - called by some as a "quaternion conjugation" one gets a conserved current\cite{welch1} duplicating electrical charge.  Thus the simplest particle spectrum totally within $U_H$ would be a single object - the up-quark (the least massive) - which interacts by the electrical force and a color exchange mediated strong force.  When one allows an interaction - a "coupling" - between the imaginaries $i$ of $U_C$ and the $I,J,K$ of $U_H$ a twofold degeneracy is created - one without the coupling and one with - and the particle spectrum now contains both the up-quark and the down-quark, the latter being more massive because of the coupling.  Finally, by coupling 1$_H$ to each of 1$_C$ and 1$_R$ the threefold family mass structure is created.  Thus the single quark achieves a six-fold degeneracy - a twofold imaginary coupling and a threefold coupling of the "reals." \\ \\
Returning to $U_R$, one should note that any particles within $U_R$ have no electromagnetic interaction - there is no "imaginary" in that sub-universe.  Thus we would expect that any matter existing solely there to be "dark."
\subsection{Mathematical Structure Summary} One must remember that a mathematical structure is being created - not a scientific theory.  The entities and relationships describing that structure must be rich enough to explain reality in the most economical manner possible.  As a solid starting point, by using all - but no more - of the suitable algebraic fields as entities and by hypothesizing a minimum number of relationships we achieve that purpose.  This structure can now be used as a road map to show how scientific theories could be constructed.  
\section{A Toy Theory Illustrating the Mathematical Structure} 
In this section an approach to a scientific theory will be developed to illustrate the use of the mathematical structure.  The theory will be not be complete and its correctness is beside the point (although the author believes that it is correct as far as it goes) because the purpose is to show how the sub-universes can be coupled.  Some features result which have an obvious physical interpretation.\\ \\
The simplest quantum mechanical problem is a linear first-order equation connecting space, time and mass (the Dirac equation) describing a "free-particle,"  which can be written in a "ring-free" (without $i \in\mathbb{C}$) fashion as:
\begin{equation}
\mathcal{D}\psi\equiv(C_x \partial_x + C_y \partial_y + C_z \partial_z + C_t \partial_t)\psi= m\psi. \label{eqn:dw}
\end{equation}
In \cite{welch}, I investigated this equation in each of $\mathbb{R}, \mathbb{C}$ and $\mathbb{H}$. In \cite{welch1} I looked specifically at $\mathbb{H}$ and showed how "color" could be explained.\\ \\
The correspondence principle (see page 890 of \cite{messiah}), requires that the solutions of ($\ref{eqn:dw}$) also satisfy the Klein-Gordon equation:
\begin{equation}
\mathcal{K}\psi\equiv(\nabla^2 - \partial_t^2)\psi =  m^2\psi , \label{eqn:KG}
\end{equation}
and thus the following conditions must hold
\begin{equation}C_{x,y,z}^2 = 1;\hspace{.1in}  C_t^2 =  -1; \hspace{.1in}\mbox{ and }\left\{C_\mu,C_\nu\right\}=C_\mu C_\nu + C_\nu C_\mu =0, \mbox{ where } \mu\neq\nu.\hspace{.1in}\mu,\nu = x,y,z,t
\end{equation} 
for whichever of $U_R, U_C$ and $U_H$ we're constructing a theory.\\ \\
Following the traditional treatment of Dirac's equation in $\mathbb{C}$ - we have in Appendix A, one representation of 4x4 matrices, the Dirac matrices, satisfying these conditions. It was shown in \cite{welch} that this representation of the Dirac matrices lead to identical results with other more traditional representations - as it should. Using these matrices, equation ($\ref{eqn:dw}$) can be written:

\begin{equation}
\mathcal{A}\psi=\left( \begin{array}{cccc}
       i\partial_t-m & 0& -i\partial_z&-i\partial_x-\partial_y\\ 
       0&i\partial_t-m&-i\partial_x+\partial_y&i\partial_z\\ 
       i\partial_z&i\partial_x+\partial_y&-i\partial_t-m&0 \\ 
       i\partial_x-\partial_y&-i\partial_z&0&-i\partial_t-m
              \end{array}\right)
              \left(\begin{array}{c} \psi_1\\ \psi_2\\ \psi_3\\ \psi_4 \end{array}\right)=0 \label{eqn:1c1}            
\end{equation} 
which, in order to have solutions, the following must hold true:
\begin{equation}
\text{det}\left|\mathcal{A}\right|=(\partial_x^2+\partial_y^2+\partial_z^2 - \partial_t^2  -m^2)^2=(\nabla^2-\partial_t^2-m^2)^2=0 \label{eqn:kg1}
\end{equation}
It should be noted that in going from ($\ref{eqn:dw}$) to ($\ref{eqn:1c1}$) the tacit assumption is made that\footnote{Remember $m$ is a \textit{c-number.} - a multiplier of $1_4$.}:
\begin{equation*}
m\psi=m(1_4) \psi=m\left( \begin{array}{cccc}
1&0&0&0\\
0&1&0&0\\
0&0&1&0\\
0&0&0&1\\
\end{array}\right)\psi
\end{equation*}whereas the only requirement to recover the Klein-Gordon equation is that
\begin{equation*}
m^2\psi=m^2(1_u)^2\psi=m^2\left( \begin{array}{cccc}
1&0&0&0\\
0&1&0&0\\
0&0&1&0\\
0&0&0&1\\
\end{array}\right)\psi
\end{equation*}
Thus any $1_u$ such that
\begin{equation*}
(1_u)^2=\left( \begin{array}{cccc}
1&0&0&0\\
0&1&0&0\\
0&0&1&0\\
0&0&0&1\\
\end{array}\right)
\end{equation*} would suffice.  It seems reasonable to confine the choices for 1$_u$ to unitary Hermitian matrices. In order to maintain linearity such that:
\begin{equation*}
\mathcal{D}\psi = (m_1 1_u+m_2 1_v)\psi\;\;\;\text{ leads to  }\\
\mathcal{K}\psi = (m_1^2+m_2^2)\psi
\end{equation*}
requires that $\{1_u,1_v\}=1_u1_v+1_v1_u = 0$, \emph{i.e.}, that they anticommute. It is well known (see page 896 of \cite{messiah}) from the study of the Dirac matrices that there are only 5 such matrices.  Two of the five have imaginary elements and, thus, in the Dirac equation would represent imaginary masses.  The three remaining matrices, in the current representation, which conveniently can be used to represent 1$_R$, 1$_C$ and 1$_H$, are:\\
\begin{equation*}
1_R=\left( \begin{array}{cccc}
 1 & 0 & 0 & 0\\
 0 & 1 & 0 & 0\\
 0 & 0 & -1 & 0\\
 0 & 0 & 0 & -1
      \end{array} \right)\;\;\;
1_C=\left( \begin{array}{cccc}
 0 & 0 & 1 & 0\\
 0 & 0 & 0 & 1\\
 1 & 0 & 0 & 0\\
 0 & 1 & 0 & 0
      \end{array} \right)\;\;\;
1_H =\left( \begin{array}{cccc}
 0 & 0 & 0 & 1\\
 0 & 0 & -1 & 0\\
 0 & -1 & 0 & 0\\
 1 & 0 & 0 & 0
      \end{array} \right)
\end{equation*}        
We will now use these - one for each of $U_R, U_C,$ and $U_H$ - in the mass term of the Dirac equation\footnote{A similar idea is used in constructing the Majorana equation as an alternative to Dirac's equation.  In that case, $\mathcal{D}\psi = m\psi_c$ where $\psi_c$ is the charge conjugate of $\psi$ and is defined as $\gamma\psi$ where $\gamma$ is a Dirac matrix.}  to "couple" sub-universes together. We will also maintain\footnote{This seems necessary - for reasons that are not clear to the author - in order to have at least one positive real mass.} 1$_4$ with a \textit{c-number} multiplier, $m$. To couple $U_R$,$U_C$ and $U_H$ we sum the relevant terms:
\begin{equation*}
\mathcal{D}\psi = (m1_4+m_R1_R+m_C1_C+m_H1_H)\psi
\end{equation*}
leading to
\begin{equation}
\mathcal{A}\psi=\left( \begin{array}{cccc}
       i\partial_t-m-m_R & 0& -i\partial_z-m_C&-i\partial_x-\partial_y-m_H\\ 
       0&i\partial_t-m-m_R&-i\partial_x+\partial_y+m_H&i\partial_z-m_C\\ 
       i\partial_z-m_C&i\partial_x+\partial_y+m_H&-i\partial_t-m+m_R&0 \\ 
       i\partial_x-\partial_y-m_H&-i\partial_z-m_C&0&-i\partial_t-m+m_R
              \end{array}\right)
              \left(\begin{array}{c} \psi_1\\ \psi_2\\ \psi_3\\ \psi_4 \end{array}\right)=0 \label{eqn:1c2}            
\end{equation} 
which, in order to have solutions, the following must hold true:
\[\text{det}\left|\mathcal{A}\right|=\left[\partial_x^2+(\partial_y+m_H)^2+\partial_z^2 - (\partial_t+im_R)^2-m^2+m_C^2 \right]^2=0\]
Note that if $m_R=m_C=m_H=0$ then ($\ref{eqn:kg1}$) is recovered.  If $m_C \ne 0$ but there is no coupling - equivalent to setting $m_R=m_H=0$ - then
\[\text{det}\left|\mathcal{A}\right|=\left[\partial_x^2+\partial_y^2+\partial_z^2 - \partial_t^2-m^2+m_C^2 \right]^2=0\]
The mass-squared, $m^2$, is reduced by an amount $m_C^2$ - having opposite signs.
Note that if $m_H \ne 0$ parity is broken\footnote{"y" is singled out here solely because of the representation chosen for the Dirac matrices.  In other representations, "x" or "z" would be involved.}
\begin{equation*} 
\left[(\partial_y+m_H)^2 \right] \ne \left[(-\partial_y+m_H)^2 \right]
\end{equation*}
Likewise if $m_R \ne 0$ then time reversal symmetry is not applicable because
\begin{equation*} 
\left[(\partial_t+im_R)^2 \right] \ne \left[(-\partial_t+im_R)^2 \right]
\end{equation*}
It is hypothesized that the mass spectrum of the leptons are due to the following couplings:
\[\begin{array}{cl}
e:& 1_4, 1_C\\
\mu:& 1_4, 1_C, 1_R\\
\tau:& 1_4, 1_C, 1_H\\
?:&1_4,1_C,1_R,1_H
\end{array}\]
and for the quark families (each split twofold by coupling between $i$ and $I,J,K$.)
\[\begin{array}{cl}
u,d:& 1_4, 1_H\\
c,s:& 1_4, 1_H, 1_R\\
t,b:& 1_4, 1_H, 1_C\\
?,?:& 1_4, 1_H,1_C,1_R
\end{array}\]
This scheme allows for the existence of a fourth family - much more massive - which has couplings among all four, $1_4, 1_R, 1_C, 1_H$. 
\section{Discussion}
A novel technique in this paper is maintaining the identity among isomorphic elements ($1_R, 1_C,1_H$) of differing mathematical fields and extracting physical features based on their distinctiveness.\\ \\  
To continue this theory one would develop representations of $i, I,J,K$ and use them as a substitute for $i$ in the term
\[(\partial_\mu -ieA_\mu)\] created by the introduction of electromagnetism in Dirac's equation.  This would couple the electromagnetic $U_C$ to the electromagnetic part of $U_H$.  The couplings between the elements of $U_C$ and $U_H$ would account for the electroweak force.  The couplings of either $U_C$ or $U_H$ to $U_R$ provide mass spectra.  The strong interaction would be introduced by relationships involving the couplings among $I,J,K$.  This point of view - again to use Tegmark's terminology - is from a bird very high up in the sky.\\ \\
The toy theory is not meant to be rigorous or complete but is meant to be a less abstract, concrete, rendition of how a non-trivial mathematical structure can be used in a less traditional manner to construct a theory.  There are enough features of this toy theory that coincide with physical reality that should persuade skeptics that there may be fruit in the approach.

\section{Acknowledments}
This work supported in part by the Social Security Administration and TIAA-CREF, contract numbers concealed to prevent identity theft.\\ \\
\noindent{
\normalsize{\textbf{Appendix A, Dirac's Matrices in $\mathbb{C}$}}}
\vspace*{.2in}
\newline
$C_x =
\left( \begin{array}{cccc}
 0 & 0 & 0 & -i\\
 0 & 0 & -i & 0\\
 0 & i & 0 & 0\\
 i & 0 & 0 & 0
      \end{array} \right)\;\;\;\; 
C_y =
\left( \begin{array}{cccc}
0 & 0 & 0 & -1\\
0 & 0 & 1 & 0\\
0 & 1 & 0 & 0\\
-1 & 0 & 0 & 0
      \end{array} \right)$  \\ \\           
\vspace*{.3in}
$C_z = 
\left( \begin{array}{cccc}
0 & 0 & -i & 0\\
0 & 0 & 0 & i\\
i & 0 & 0 & 0\\
0 & -i & 0 & 0
       \end{array} \right)\;\;\;\;
C_t =
\left( \begin{array}{cccc}
i & 0 & 0 & 0 \\
0 & i & 0 & 0 \\
0 & 0 & -i & 0\\
0 & 0 & 0 & -i 
\end{array} \right)$  
\bibliography{quaternion}
\bibliographystyle{unsrt}
\end{document}